\documentstyle[sprocl]{article}
\def\MS{{{\hbox{MS}}}}

\def\MSf{{\tiny{\hbox{MS}}}}

\def\pd{\partial}

\def\p{\phi}
\def\L{\Lambda}
\def\La{\L}
\def\m{\mu}
\def\l{\lambda}
\def\la{\l}

\def\h{\hbox{${1\over2}$}}

\def\bib{\bibitem}

\def\be{\beta}

\def\ka{\kappa}

\def\e{\epsilon}
\def\d{\pd}
\def\k{\kappa}
\def\beq{\begin{equation}}
\def\eeq{\end{equation}}
\def\bed{\begin{displaymath}}
\def\eed{\end{displaymath}}
\def\beqq{\begin{eqnarray}}
\def\eeqq{\end{eqnarray}}
\def\bedd{\begin{eqnarray*}}
\def\eedd{\end{eqnarray*}}

\begin{document}

\title{MULTISCALE RENORMALIZATION AND DECOUPLING \footnote{
Work done in collaboration with C. Wiesendanger, email:
wie@stp.dias.ie}}

\author{Christopher Ford \footnote{
email: Ford@hpfs1.physik.uni-jena.de}}
\address{Theoretisch-Physikalisches Institut\\
Universit\"at Jena, Fr\"obelsteig 1\\
D-07743 Jena, Germany}

\maketitle
\abstracts{The standard $\MS$ renormalization
prescription is inadequate for dealing with multiscale problems.
To illustrate this, we consider the computation of the effective potential
in the Higgs-Yukawa model.
It is argued that the most natural way to deal with this problem is to
 introduce a 2-scale renormalization group. We review various
ways of implementing this idea and consider to what extent
they fit in with the notion of heavy particle decoupling.}

\section{Introduction}

Let us consider a very simple problem in perturbative quantum field theory,
the computation of the effective potential in the four-dimensional
Higgs-Yukawa model defined by the Lagrangian
\beq {\cal L}=\frac{1}{2}(\d_\m\p)^2-\frac{1}{2}m^2\p^2
-\frac{\l}{24}\p^4+\bar{\psi}i\d\!\!\!/\psi+g\bar{\psi}\p\psi+\L.
\eeq
Here $\L$ is a ``cosmological constant'' term which enters non-trivially
into the renormalization group equation for the effective potential \cite{kast}
(see also Christian Wiesendanger's talk). It is well known how to perform
a loopwise perturbative expansion of the effective potential \cite{jac},
$V(\p)=V^{(0)}(\p)+\hbar V^{(1)}(\p)+\hbar^2V^{(2)}(\p)+...\,
$.
 Using dimensional
regularization together with (modified) minimal subtraction gives
\beqq
V^{(0)}(\p)&=&\frac{\l}{24}\p^4+\frac{1}{2}m^2\p^2+\L,\\
V^{(1)}(\p)&=&\frac{(m^2+\h\l\p^2)^2}{4(4\pi)^2}\left[
\log\frac{m^2+\h\l\p^2}{\m^2}-\frac{3}{2}\right]
-\frac{g^4\p^4}{(4\pi)^2}\left[
\log \frac{g^2\p^2}{\m^2}-\frac{3}{2}\right]. \nonumber
\eeqq
Notice the logarithmic terms in the one-loop potential. We have
a $\log \frac{m^2+\h\l\p^2}{\m^2}$ due to the ``Higgs'' loop
and $\log \frac{g^2\p^2}{\m^2}$ associated with the fermionic
contribution to the one-loop potential. The two-loop potential,
$V^{(2)}$,  is quadratic in these logarithms, and in general the
$n$-loop potential is a $n$th order polynomial in the two logarithms.
Thus, for believable perturbation theory one must not only
have ``small'' couplings $\hbar \lambda$, $\hbar g^2$, but the two
logarithms must also be small. As was explained a long time ago
by Coleman and Weinberg (CW) \cite{cw}  one must make a ($\p$-dependent) choice
of $\m$ such that the logarithms are not too large. To relate the 
renormalized parameters at different scales one uses the renormalization group
(RG).
The CW procedure of RG ``improving'' the potential is equivalent to
a resummation of the large logarithms in the perturbation series.

However, it is not too difficult to see that if $m^2+\h\l\p^2>>g^2\p^2$
(the heavy Higgs case) or $g^2\p^2>>m^2+\h\l\p^2$ (a heavy fermion)
\sl there is no choice of $\m$ that will simultaneously render
both logarithms small.\rm\, Thus, we are only able to implement the CW
method when $m^2+\l\p^2\sim g^2\p^2$, ie. when we have essentially
a one-scale problem.

\section{ Multiscale renormalization}
 
We have seen that in the $\MS$ scheme the RG equation is not powerful
enough to deal with multiscale problems. In a single scale problem
one is able to track the relevant scale with the $\MS$ RG scale $\mu$,
whereas in a 2-scale model it is not possible to track two scales
with a \sl single \rm RG scale.
In order to deal with the 2-scale case it seems natural to seek
a \sl 2-scale version of \rm $\MS$. This 2-scale scheme, should
be as similar to $\MS$ as possible, but with \sl two \rm RG
scales $\ka_1$ and $\ka_2$ instead of one $\MS$ scale $\m$.
Here $\ka_1$ and $\ka_2$ should ``track'' the Higgs and fermionic scale,
respectively. Using such a 2-scale scheme one should be able
to sum up the two logarithms in the perturbation series.
For attempts to deal with this problem while retaining a single scale RG
see refs. \cite{onescale}.

How do we define a 2-scale subtraction scheme? In fact, a multiscale
renormalization scheme has already been proposed by Einhorn and
Jones (EJ) \cite{ej}  which has some of the properties we seek.
To motivate their idea, let us look at the bare Lagrangian for
our Higgs-Yukawa problem written in terms of the usual
$\MS$ renormalized parameters.
\beqq
{\cal L}_{\hbox{Bare}}&=&\frac{1}{2}Z_\p(\d_\m\p)^2-
\frac{1}{2}Z_\p Z_{m^2} m^2 \p^2-
\frac{1}{24}Z_\p^2 Z_\l {\mu}^{\e} \l \p^4\nonumber\\
&&
+
Z_\psi\bar{\psi}i\d\!\!\!/\psi+
Z_\psi Z_g {Z_\p}^{\h} {\m}^{\h \e} g\bar{\psi}\p\psi+
\L+(Z_\L-1){\mu}^{-\e}m^4\l^{-1}, \label{3}
\eeqq
where $\e=4-d$ is the dimensional continuation parameter,
and all the $Z_.$ factors have the form
$Z_.=1+\hbox{pole terms only}$. Notice that the $\MS$ RG scale
$\mu$ enters eqn. (\ref{3}) in three places.
The EJ idea was simply to replace the three occurrences of
$\m$ in (\ref{3})  with three \sl independent \rm\,  RG scales
$\k_1$, $\ka_2$, $\ka_3$, so that
\beq
\l_{\hbox{Bare}}=\k_1^{\e}Z_\l \l,
\quad
g_{\hbox{Bare}}=\k_2^{\h \e} Z_g g,\quad
\L_{\hbox{Bare}}=\La+\k_3^{-\e}(Z_\L-1) m^4\l^{-1}.
\eeq
As in standard $\MS$, the $Z_.$ factors are defined
by the requirement that the effective action is finite
when written in terms of the renormalized parameters
and the restriction that the $Z_.$ factors have
the form $Z_.=1+\hbox{pole terms only}$. Note that the $Z_.$
factors will \sl not \rm be the same as the $\MS$ $Z_.$ factors
(except where $\k_1=\k_2=\k_3$). In the EJ scheme, the $Z_.$'s will
contain logarithms of the RG scale ratios.

We now have three separate RG equations associated with the independent
variations of the three RG scales. We also have three sets of beta
functions 
\beq \label{5} {}_i\beta_\la=\k_i \frac{d}{d\k_i}\la\quad
i=1,2,3.\eeq
and similarly for the other parameters. 
It is straightforward to compute the one-loop beta functions in the EJ
scheme:
$$
{}_2\be_g=\frac{5\hbar g^3}{3(4\pi)^2},\quad {}_1\be_g={}_3\beta_g=0, $$
\beq\label{6}
{}_1\be_\l=\frac{\hbar}{(4\pi)^2}(3\l^2+48g^4),\quad
{}_2\be_\l=\frac{\hbar}{(4\pi)^2}(8\l g^2-96 g^4),\quad
{}_3\be_\l=0.
\eeq
One may be tempted now to turn these one-loop RG functions into
running couplings via eqs. (\ref{5}). However, if one were to compute
the two-loop beta functions, one would find terms proportional
to $\log \frac{\k_1}{\k_2}$, and in general the $n$-loop RG functions
contain $\log^{n-1}\frac{\k_1}{\k_2}$ terms(as well as lower powers of
the logarithm). Therefore, unlike in standard $\MS$ we \sl
cannot trust the perturbative RG functions. \rm
So if we still wish to use the EJ scheme we must \sl somehow \rm
perform a large logarithms expansion on the \sl beta functions 
themselves. \rm

Another problem with the EJ prescription is that although it has
two RG scales (three if you include $\k_3$ which is only relevant
to the running cosmological constant) $\k_1$ and $\k_2$, they
do not seem to ``track'' the Higgs and fermionic scales,
respectively. If such a tracking were present we would expect the one-loop
beta function for $\l$ to have the form
\beq\label{7}
{}_1\beta_\l=\frac{3\hbar\l^2}{(4\pi)^2},\quad 
{}_2\beta_\l=\frac{\hbar}{(4\pi)^2}(8\l g^2 -48g^4).
\eeq
That is the contributions to ${}_1\beta_\l$ and ${}_2\beta_\l$
can be identified with contributions from the
Higgs and fermion loop, respectively.
So although the EJ proposal is very interesting, it is not quite
what we were looking for. However, it may still be that some (possibly
quite simple) modification of the EJ scheme does the job.
Another possibility would be to construct a multiscale version of the 
Callan-Symanzik equation \cite{ni}.

Although we are unable to define such a modified EJ scheme we can exploit
the fact that \sl any \rm multiscale scheme must be related to the
standard $\MS$ prescription by a \sl finite \rm renormalization.
That is if we have a scheme with two RG scales $\k_1$ and $\k_2$
then we must \footnote{
We assume that  the transformation has a trivial
dependence on $m^2$.}
have
\beqq
g_\MSf&=&F_g(g,\l;\k_1,\k_2,\m)\\ \nonumber
\l_\MSf&=&F_\l(g,\l;\k_1,\k_2,\m)\\ \nonumber
m^2_\MSf&=&m^2F_{m^2}(g,\l; \k_1,\k_2, \m),
\eeqq
with similar relations for $\L$, $\p$ and $\psi$.
Here, the $\MS$ parameters $g_\MSf$, $\l_\MSf$, etc.
at scale $\m$ may be regarded as ``bare'' ones as opposed
to the new ``renormalized'' 2-scale parameters $g$, $\l$, etc.
The (finite) $F_.$ functions are chosen so that:

i) The effective action $\Gamma$, when expressed in terms of the new
 parameters should be independent of the $\MS$ scale $\m$.

ii) When $\k_1=\k_2$ the 2-scale scheme should coincide with $\MS$
at that scale.

There are an infinite number of 2- scale schemes (ie. $F_.$
functions) satisfying
conditions i) and ii). Each of these schemes will have  different
 beta functions.
Let us now restrict ourselves to schemes with the correct one-loop
tracking behaviour. That is we assume that the one-loop beta functions
for $\l$ are as in eqn. (\ref{7}). This tracking assumption also fixes
the one-loop RG functions for $m^2$, $\L$, $\p$ and $\psi$. The tracking
assumption does \sl not \rm fix the one-loop beta functions for $g$;
all we can say is that
\beq {}_1\beta_g=\frac{\e_1 \hbar g^3}{(4\pi)^2},\quad
{}_2\beta_g=\frac{\e_2 \hbar g^3}{(4\pi)^2},
\quad\hbox{where }
\e_1+\e_2=\frac{5}{3}.
\eeq
A problem with the EJ scheme was the occurrence of logarithmic terms in the
higher loop RG functions. Is it possible to devise a 2-scale scheme
where the beta functions have no such logarithms?
The answer to this question is no, and so anyone wishing to generalize
the EJ scheme must face the problem of resumming logs in the beta functions
themselves. To see this consider the two RG equation for the effective
 potential
\beq
{\cal D}_iV=0,\quad
{\cal D}_i=\k_i\frac{\d}{\d k_i}+{}_i\beta_g
\frac{\d}{\d g}+{}_i\beta_\l\frac{\d}{\d\l}+
{}_i\beta_{m^2}\frac{\d}{\d m^2}+{}_i\beta_\L\frac{\d}
{\d\L}-{}_i\beta_\p\p\frac{\d}{\d\p},
\eeq
where $i=1,2$ and the summation convention was not used in the last
 equation. We have the integrability condition
\beq\label{11}
[{\cal D}_1,{\cal D}_2]=0.
\eeq
The point is that the absence of logs in the RG functions, ie.
$[\k_i\d/\d\k_i,{\cal D}_j]=0$ is \sl incompatible \rm with the integrability
condition eqn. (\ref{11}).

However, it is still possible to arrange for \sl one of \rm
the two sets of beta 
functions to be independent of $\k_1/\k_2$, eg.
we can take the first set of beta functions to be independent of
$\k_1/\k_2$. Alternatively, we can take the second set of RG functions 
(tracking the fermionic scale) to be independent of $\k_1/\k_2$.
Whichever of these prescriptions we adopt, we can then use the integrability
condition to resum the logarithms in the other set of beta functions.
Some detailed calculations of this type (though in a different
model) have been given in ref. \cite{fowi}.

\section{Decoupling and Conclusions}

We have argued that it is possible to construct a 2-scale scheme
with appropriate tracking at one-loop  where one of the two sets
of RG functions is independent of the RG scales. Let us consider the
case where we require that the first set of beta functions (tracking
the Higgs scale) is independent of $\k_1/\k_2$. Then at leading order
the first set of beta functions are
\beq\label{dec}
{}_1\beta_\l=\frac{3\hbar\l^2}{(4\pi)^2},\quad
{}_1\beta_{m^2}=\frac{\hbar\l m^2}{(4\pi)^2},\quad
{}_1\beta_\L=\frac{\hbar m^4}{2(4\pi)^2},\quad
{}_1\beta_\p=0,\quad
{}_1\beta_g=\frac{\e_1\hbar g^4}{(4\pi)^2}.
\eeq
Clearly, these beta functions are just the usual one-loop beta functions
for pure $\p^4$ theory (provided we make the choice $\e_1=0$).
The second set of beta functions depend on $\k_1/\k_2$; this dependence
can be computed via the integrability condition (\ref{11}).
Now if the fermion is much heavier than the Higgs scale
we would \sl expect \rm\, that the beta functions for the low energy
theory would be exactly those given by eqn. (\ref{dec}), since
in this case we expect to observe a decoupling \cite{decouple}
of the heavy fermion. Thus, it seems that the condition that
the first set of beta functions is independent of the RG scales
is appropriate for the heavy fermion case
($g^2\p^2>>m^2+\h\l\p^2$). Similarly, one can argue that the alternative
possibility of requiring that the second set of beta functions is
independent of $\k_1/\k_2$ is suited to the heavy Higgs case
($m^2+\h\l\p^2>>g^2\p^2$).
Note that although in the previous section we were unable to fix the values
of $\e_1$ and $\e_2$ via a one-loop tracking condition, an explicit
calculation shows that the final improved potential only depends
on $\e_1$ and $\e_2$ through the combination $\e_1+\e_2=
\frac{5}{3}$.

 From decoupling arguments we
 expect the case where the first set
 of
beta functions has no logarithms is suited to the heavy fermion
case. We believe that this approach would correctly interpolate between
 the heavy fermion case and the single scale regime 
($m^2+\h\l\p^2\sim g^2\p^2)$, since by construction our two
scale scheme collapses to $\MS$ for $\k_1=\k_2$.
There is no reason to expect that this
prescription would be valid in the heavy Higgs case (for this we would
use the alternative possibility where the second set of beta functions
is independent of the RG scales). Thus, it seems that we can deal with
 both
the heavy fermion and heavy Higgs problem, but these require a separate
treatment. It remains an open question whether it is possible
to devise a simple EJ type scheme which can interpolate
all the way from the heavy fermion to the heavy Higgs sector.

\section*{Acknowledgments}
I am grateful to the organizers of RG96 for giving me the opportunity
to speak at this conference. Thanks also to 
Denjoe O'Connor and Chris Stephens for explaining some of their ideas  on
  multiscale
problems.

\end{document}